\documentclass[aps,pra,reprint,superscriptaddress,
 amsmath,amssymb]{revtex4-1}
\usepackage[utf8x]{inputenc}
\usepackage{systeme}
\usepackage{color}
\usepackage{setspace}
\usepackage{float}
\usepackage{multirow}
\usepackage{amssymb}
\usepackage[linesnumbered,boxed]{algorithm2e}
\usepackage{amsmath,tabu}  
\usepackage{bbold}
\usepackage{esint}
\usepackage{graphicx}
\usepackage{dcolumn}
\usepackage{bm}

\usepackage{hyperref}
\hypersetup{colorlinks=true,       
			linkcolor=black,        
			citecolor=blue,        
			filecolor=magenta,     
			urlcolor=blue}

\begin{document}


\title{Strong quantum correlations of light emitted by a single atom in free space}

\author{D. Goncalves}
\email{daniel.goncalves@icfo.es}
\affiliation{ICFO-Institut de Ciencies Fotoniques, The Barcelona Institute of Science and Technology, 08860 Castelldefels (Barcelona), Spain}
\author{M. W. Mitchell}
\affiliation{ICFO-Institut de Ciencies Fotoniques, The Barcelona Institute of Science and Technology, 08860 Castelldefels (Barcelona), Spain}
\affiliation{ICREA-Institució Catalana de Recerca i Estudis Avançats, 08015 Barcelona, Spain.}
\author{D. E. Chang} 
\affiliation{ICFO-Institut de Ciencies Fotoniques, The Barcelona Institute of Science and Technology, 08860 Castelldefels (Barcelona), Spain}
\affiliation{ICREA-Institució Catalana de Recerca i Estudis Avançats, 08015 Barcelona, Spain.}

\date{\today}

\begin{abstract}
We present a novel approach to engineer the photon correlations emerging from the interference between an input field and the field scattered by a single atom in free space. Nominally, the inefficient atom-light coupling causes the quantum correlations to be dominated by the input field alone. To overcome this issue, we propose the use of separate pump and probe beams, where the former increases the atomic emission to be comparable to the probe. Examining the second-order correlation function $g^{(2)}(\tau)$ of the total field in the probe direction, we find that the addition of the pump formally plays the same role as increasing the coupling efficiency. We show that one can tune the correlation function $g^{(2)}(0)$ from zero (perfect anti-bunching) to infinite (extreme bunching) by a proper choice of pump amplitude. We further elucidate the origin of these correlations in terms of the transient atomic state following the detection of a photon.
\end{abstract}

\maketitle
\section{Introduction}
Despite its apparent simplicity, the interference between an optical field incident on a single atom and the field scattered by such a quantum nonlinear element gives rise to a wealth of phenomena. For example, it can produce strong bunching or anti-bunching of the re-radiated fields \cite{ref:single_atom, ref:1D_waveguide, ref:Quantum_aperture}, or yield a ``quantum aperture" for light with a rich spatial structure of quantum photon correlations \cite{ref:Quantum_aperture}. It can also give rise to non-trivial stimulated emission statistics at the quantum level \cite{ref:Fan}, induce multi-photon bound states in scattering \cite{ref:multi_photon_boundstate_1,ref:multi_photon_boundstate_2} or produce photon number-dependent propagation delays \cite{ref:Giuseppe}. To realize the majority of these effects, however, it is necessary to achieve highly efficient interactions between individual photons and the atom, such that the incident light and re-scattered fields become comparable in strength, leading to a large interference.

Unfortunately, the single photon-atom interaction efficiency is intrinsically weak in free space. For a single incident photon focused to an area $A$ and with a wavelength $\lambda$ resonant with the transition of a two-level atom, the interaction efficiency scales $\sim\lambda^2/A$ \cite{ref:Quantum_aperture}. However, the diffraction limit ($A\gtrsim \lambda^2$) and subtleties associated with tight focusing \cite{ref:Focusing_single_atom_Kurtsiefer,ref:VanEnk_focusing} constrain this interaction probability to be about $\sim10\%$ in current experiments \cite{ref:Kurt_experiment,ref:Historical_context,ref:Kurtsiefer}. For such low coupling efficiencies, the total field is instead dominated by the (classical) input field. Thus far, the only routes to approach unity coupling efficiency have involved either high-finesse cavities \cite{ref:cavity_strong_coupling,ref:Rempe_reiserer} or waveguides with strong field confinement \cite{ref:Waveguide_2,ref:Waveguide_1}.
 \begin{figure}
 \centering
    \includegraphics[width=0.72\linewidth]{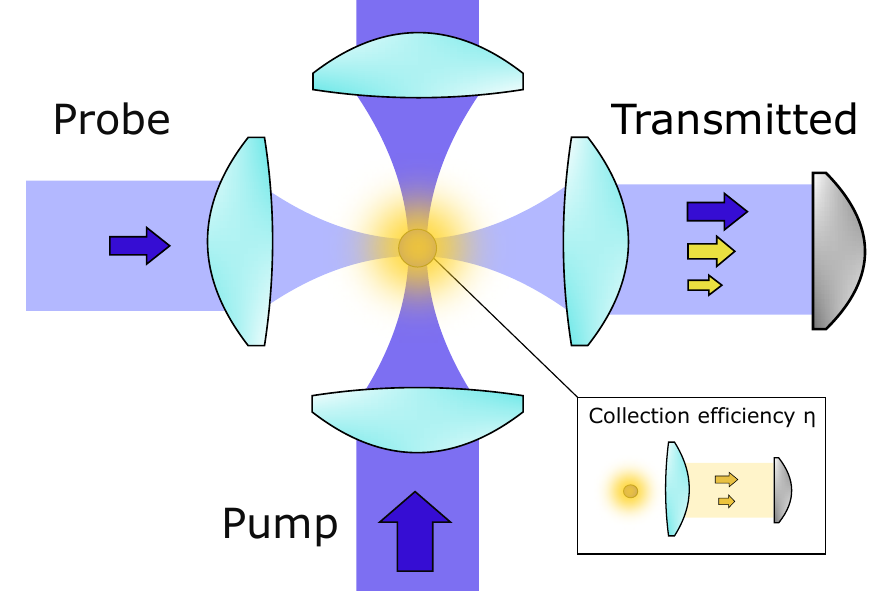}
  \caption{Conceptual scheme of the proposed technique. Two coherent, continuous beams (pump and probe) illuminate resonantly a single atom in a ``Maltese cross" configuration. Along the transmitted direction of the probe beam, the total field consists of the coherent sum of the incident probe field (small blue arrow) and the quantum field re-scattered by the atom, which contains contributions from both the probe and pump beams (small and large yellow arrows, respectively). A sufficiently large pump beam allows the re-scattered field to be comparable to the incident probe in intensity, despite a low collection efficiency $\eta$ of emitted photons in the transmitted direction. This enables strong quantum correlations to emerge in the  collected field.}
    \vspace{-1em}
  \label{fig:setup}
\end{figure}

In this work, we present a novel approach to observe and manipulate quantum interference effects between an incoming field and a single free-space atom, even in the low coupling regime. The key idea is to use two separate pump and probe fields propagating in spatial modes that only significantly overlap at the atomic position (Fig. 1) \cite{ref:Morgan}. Then, the total transmitted light collected in the probe direction consists of the input field and the quantum re-scattered field, which contains contributions from both pump and probe beams. By tuning the pump strength, the total scattered field can become comparable to the probe amplitude, allowing for rich correlations to emerge. In particular, we show that as far as the transmitted intensity and the second-order field correlation function $g^{(2)}(\tau)$, the addition of the pump formally plays the same role as increasing the coupling efficiency. This can be used to tune between fully anti-bunched ($g^{(2)}(0)\rightarrow 0$) and extremely bunched ($g^{(2)}(0)\rightarrow \infty$) second-order correlation functions of the total field. We also provide an interpretation of the physical origin of these correlations from the atomic state perspective.

\section{Theoretical model}
 Generally, our goal will be to calculate the spatio-temporal properties and correlations of a quantum field, as it propagates and interacts with an atom as in Fig. \ref{fig:setup}, for which we briefly present our theoretical formalism here. First, we consider the dynamics of an ideal, two-level atom with ground state $|g\rangle$ and excited state $|e\rangle$ driven by a resonant input field with Rabi frequency $\Omega$ representing the combination of classical probe and pump input beams. The time evolution of the quantum atomic state $\hat{\rho}$ obeys the master equation
\begin{equation}
    \centering
        \dot{\hat{\rho}}=-\frac{i}{\hbar}[\hat{H},\hat{\rho}]+\frac{\Gamma_0}{2}\left(2\hat{\sigma}^{ge}\hat{\rho}\hat{\sigma}^{eg}-\hat{\sigma}^{ee}\hat{\rho}-\hat{\rho}\hat{\sigma}^{ee}\right),
        \label{eq:Master_equation}
\end{equation}
where $\Gamma_0$ is the free-space spontaneous emission rate and $\hat{\sigma}^{ab}=| a\rangle \langle b|$ are atomic operators with $\{a,b\}\in\{e,g\}$. The Hamiltonian $\hat{H}$ from Eq. ($\ref{eq:Master_equation}$) contains the interaction term between the atom and a resonant, quantum driving field. Explicitly, it takes the form $\hat{H}=-\hbar \left(\Omega\hat{\sigma}^{eg}+h.c.\right)$ in the laser's rotating frame. Whereas the spontaneous emission term of Eq. (\ref{eq:Master_equation}) implicitly encodes the loss of atomic excitation in the form of radiated photons, the explicit spatio-temporal properties of this re-scattered field can be found through the input-output relation \cite{ref:input_output_2,ref:input_output_1}
\begin{equation}
    \hat{\textbf{E}}_{\text{out}}(\textbf{r})=\hat{\textbf{E}}_{\text{in}}(\textbf{r})+\mu_0\omega_{ge}^2\textbf{G}_0(\textbf{r},\textbf{r}_\text{a}, \omega_{ge})\cdot \textbf{d}_{ge}\ \hat{\sigma}^{ge},
    \label{eq:Input_output_not_projected}
\end{equation}
where $\hat{\textbf{E}}_{\text{in}}(\textbf{r})$ is the input field operator and $\textbf{r}_\text{a}$ the position of the atom. The quantity $\textbf{d}_{ge}$ is the dipole matrix element associated to the transition $|g\rangle \leftrightarrow |e\rangle$ with frequency $\omega_{ge}$ and is connected to $\Gamma_0$ through the relation $\Gamma_0=|\textbf{d}_{ge}|^2\omega_{ge}^3/(3\pi\hbar \epsilon_0c^3)$ \cite{ref:textbook}. $\textbf{G}_0(\textbf{r},\textbf{r}_\text{a},\omega_{ge})$ is the Green's function in free space, which encodes the field emitted by a point-like dipole source and satisfies the electromagnetic wave equation \cite{ref:Jackson}
\begin{equation}
    \left[\textbf{$\nabla$} \times \textbf{$\nabla$}\times-\frac{\omega_{ge}^2}{c^2}\mathbb{1}\right]\textbf{G}_0(\textbf{r},\textbf{r}',\omega_{ge})=\delta (\textbf{r}-\textbf{r}')\mathbb{1}.
\end{equation}
Its explicit form is given by
\begin{multline}
    \textbf{G}_0(\textbf{r},\textbf{r}', \omega_{ge})=\frac{e^{ik_0R}}{4\pi R}\Bigg{[}\left(1+\frac{i}{k_0R}-\frac{1}{k_0^2R^2}\right)\mathbb{1}-\\
    -\left(1+\frac{3i}{k_0R}-\frac{3}{k_0^2R^2}\right)\frac{\textbf{R}\otimes \textbf{R}}{R^2}\Bigg{]},
\end{multline}
where $R=|\textbf{r}-\textbf{r}'|$ and $k_0=\omega_{ge}/c$. The tensor nature of the Green's function accounts for the vectorial nature of both the dipole source orientation and the emitted field. Eq. (\ref{eq:Input_output_not_projected}) states that the total field can be decomposed into an incident field and a field re-scattered by the atom, an idea well-known in classical optics. Importantly, however, such a relation also holds true as an operator equation. In particular, it enables the quantum correlations of the output field $\hat{\textbf{E}}_{\text{out}}(\textbf{r})$ to be calculated in terms of the input field and atomic state (the latter being encoded in the solution to Eq. (\ref{eq:Master_equation})). \\

Although Eq. (\ref{eq:Input_output_not_projected}) is formally true, measuring the fields at a single point $\textbf{r}$ is not typical from an experimental perspective. One commonly-used detection modality is to collect the transmitted light through the atom with an optical imaging system, which projects the total field into a certain spatial mode. To provide a specific example, in Fig. \ref{fig:setup}, the probe field might consist of a focused Gaussian beam, while in the transmitted direction, light is collected back into the same Gaussian mode. Assuming that the input and detection spatial modes $\textbf{E}_{\text{det}}(\textbf{r})$ are the same, one can project the operator $\hat{\textbf{E}}_{\text{out}}(\textbf{r})$ from Eq. (\ref{eq:Input_output_not_projected}) into this preferred mode, to obtain (see Appendix)\cite{ref:Marcos_paper} 
\begin{equation}
    \hat{E}_{\text{det}}=\hat{E}_{\text{in,det}}+i\sqrt{\eta\Gamma_0}\ \hat{\sigma}^{ge}
    \label{eq:Input_output_projected},
\end{equation}
where $\hat{E}_{\text{in,det}}$ is the input field in the detection mode. Here for convenience, we have re-scaled the field operators, so that $\langle\hat{E}_{\text{det}}^\dagger \hat{E}_{\text{det}}\rangle$ represents the total number of photons per unit time. The projection of the atomic scattered field into the detection mode is encoded in a single parameter, $\eta$, which physically describes the collection efficiency of a single emitted photon into this mode. By time reversal symmetry, it also provides the interaction efficiency between the atom and an incoming resonant photon \cite{ref:time_reversed, ref:retrieval_efficiency}.\\

As an example, we can consider a Gaussian beam of waist $w$, which is focused onto the atom and whose polarization is aligned along with the dipole matrix element. The spatial mode is given by $E_{\text{det}}(\rho,z=0)=E_0e^{-\rho^2/w^2}$ in the paraxial approximation, with $\rho$ being the radial distance from the center axis of the beam. By conveniently choosing the atomic position to coincide with the focal point, we find the well-known result that the interaction efficiency is $\eta\approx3\lambda^2/8\pi^2w^2$ (see Appendix)\cite{ref:Quantum_aperture}.

One could go beyond the paraxial approximation to account for different corrections, such as the distortion of the polarization due to tight focusing of the beams \cite{ref:VanEnk_focusing} or effects related to the focusing lens \cite{ref:Focusing_single_atom_Kurtsiefer}. However, for low values of $\eta$, i.e. low focusing of the fields, the paraxial approximation agrees very well with full vector solutions \cite{ref:paraxial_approx}. In any case, our results below will be given in terms of $\eta$ to be as general as possible, and thus are independent of such considerations.

\section{Pump-probe scheme}
 Having introduced a rather general formalism to calculate the properties of the quantum field, we now apply it specifically to the pump-probe scenario introduced earlier. As a first step, we explore how the pump modifies the transmission of light in the probe mode.
 \subsection{Transmission in the collection direction}
Following the discussion from Sec. II, our strategy will be to first obtain the atomic state driven by the total field (Eq. \ref{eq:Master_equation}) and use it to construct field correlations according to Eq. (\ref{eq:Input_output_projected}). Although the single-atom density matrix can be readily solved in general, here we focus on the weak driving regime where $|\Omega|\ll\Gamma_0$, which already contains the interesting physics. In this limit, to lowest order, the steady state solution of Eq. (1) is $\hat{\rho}_{eg}\approx 2i\Omega/\Gamma_0$ for the atomic coherence and $\hat{\rho}_{ee}\approx |\hat{\rho}_{eg}|^2$ for the population in the excited state. Their specific values depend on the Rabi frequency of the total field driving the atom, which is the admixture of the pump and the probe contributions, $\Omega= \Omega_{\text{pump}}+\Omega_{\text{probe}}$. Since the physics will also depend on the relative phase $\phi$ between the probe and the pump beams, we take $\Omega_{\text{probe}}>0$ to be positive and real, while allowing $\Omega_{\text{pump}}=|\Omega_{\text{pump}}|e^{i\phi}$ to be complex.\\

Once the atomic state is known, we calculate the transmission of light in the probe mode. We assume that the pump beam is in a completely orthogonal mode to the probe, such that its contribution to the detected field in Eq. (\ref{eq:Input_output_projected}) only comes via the atomic scattered field. We define the transmission coefficient $T$ as the ratio between intensities seen in the detection mode and the probe input, i.e $T=\langle \hat{E}_{\text{det}}^\dagger\hat{E}_{\text{det}}\rangle /\langle \hat{E}_{\text{in,det}}^\dagger \hat{E}_{\text{in,det}}\rangle $. It can be shown that for normally ordered correlation functions and coherent state inputs, the input field operator $\hat{E}_{\text{in,det}}$ in Eq. (4) can be replaced by the square root of its corresponding coherent state photon flux $\sqrt{\Phi_\text{p}}$ \cite{ref:Substitution}, according to our normalization for the operators. Substituting Eq. (\ref{eq:Input_output_projected}) in the definition for the transmission coefficient, we find
 \begin{equation}
     T=1-2\sqrt{\eta\Gamma_0}\frac{\text{Im}\left\{\hat{\rho}_{eg}\right\}}{\sqrt{\Phi_\text{p}}}+\eta\Gamma_0\frac{ \hat{\rho}_{ee}}{\Phi_\text{p}},
     \label{eq:Transmission}
 \end{equation}
which explicitly depends on the atomic coherence and population in the excited state, for which we substitute the aforementioned values in the weak driving regime. The resulting expression contains both $\Omega_{\text{probe}}$ and the probe photon flux $\Phi_\text{p}$, which are related through $\eta\Phi_\text{p}=|\Omega_{\text{probe}}|^2/\Gamma_0$ (see Appendix). With this, it is straightforward to show that
 \begin{equation}
     T=\left|1-2\Lambda\right|^2,
     \label{eq:Transmission_clean}
 \end{equation}
 where $\Lambda=\eta(\Omega/\Omega_{\text{probe}})$ is an effectively enhanced coupling parameter that depends on three quantities: the ratio $|\Omega_{\text{pump}}/\Omega_{\text{probe}}|$, the pump-probe relative phase $\phi$ and the free-space atom-photon coupling efficiency $\eta$. \\
 
We now discuss the implications of Eq. (\ref{eq:Transmission_clean}) starting with the case without the pump beam, where the transmission takes the well-known value $T=|1-2\eta|^2$ for single-atom resonant transmission \cite{ref:Blatt,ref:Kurtsiefer,ref:1D_waveguide}. For small coupling efficiencies, the light is only weakly attenuated as the atom can only scatter a small fraction of the incoming photons. If the probe input mode occupies the half-space (of solid angle), and moreover matches the atomic dipole emission pattern, the coupling efficiency would reach $\eta = 1/2$, achieving perfect attenuation $T=0$ (along with perfect reflection $R=1$ in the backward direction). Now, when we introduce the pump, interestingly we see that for $\phi=0$, the parameter $\Lambda$ is equivalent to a renormalized coupling efficiency. In particular, even if $\eta$ is small, one can tune $\Omega_{\text{pump}}$ to obtain $\Lambda=1/2$, achieving the previous perfect ``attenuation" $T=0$. In this case, the atom elastically and coherently scatters photons from the larger pump beam into the collection direction, with an amplitude and phase that can cancel the incoming probe (Fig. \ref{fig:setup}). Of course, the addition of the pump can also increase the total number of photons scattered in the collection direction to exceed the number of input probe photons, causing $T>1$. \\
 
\begin{figure}[h!]
    \includegraphics[width=1\linewidth]{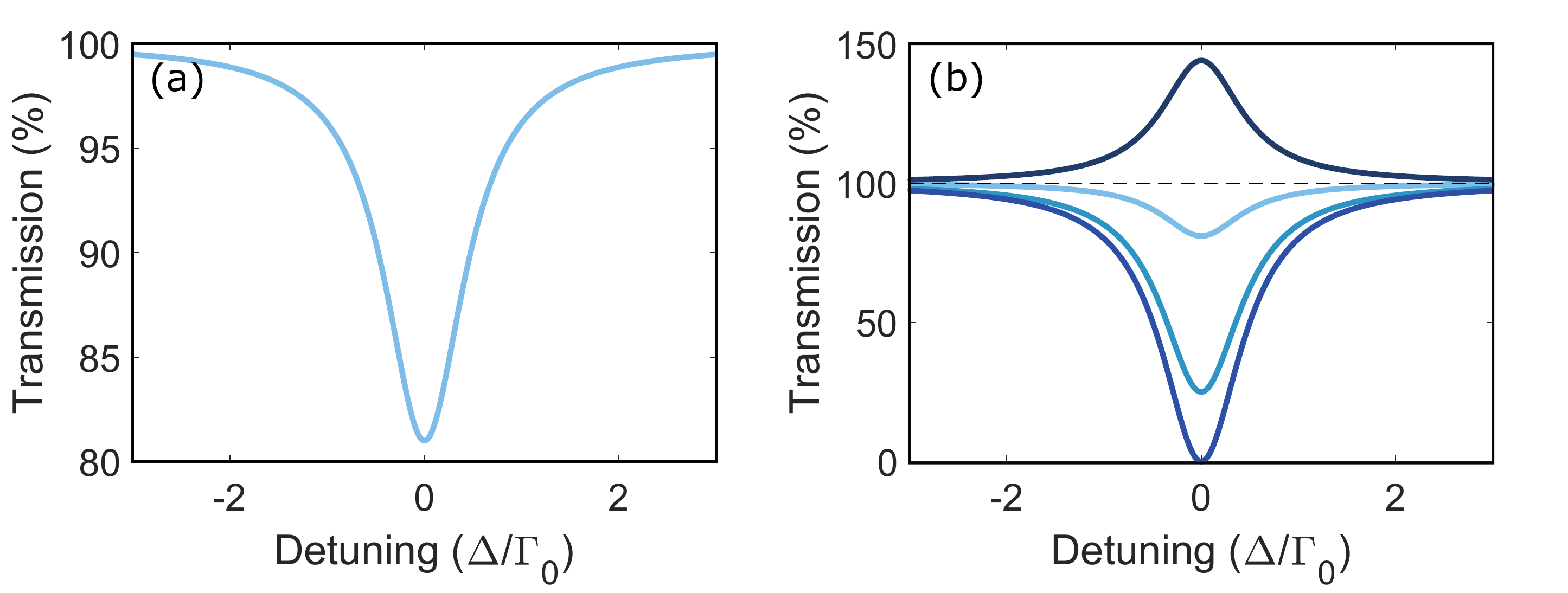}
    \caption{Transmission spectra of a weak, coherent probe beam as a function of the laser detuning $\Delta$ with respect to the atomic transition, in units of the free-space decay rate $\Gamma_0$. (a) Transmission spectrum without a pump beam. We take a realistic coupling efficiency of $\eta=0.05$, consistent with typical experimental values.  (b) Same as (a) but now with an additional pump beam, with pump-probe relative phase $\phi=0$. The pump amplitude is tuned to obtain different effective coupling efficiencies $\Lambda=\{\eta, 0.25, 0.5,1.1\} $ (colors from light to dark blue). We note a total extinction of the resonant transmission for $\Lambda=0.5$ and values of transmission higher than 100\% for $\Lambda>1$.}
    \label{fig:Transmission}
\end{figure}

 The ability to tune the transmission coefficient is illustrated in Fig. \ref{fig:Transmission}. Here, we plot the spectrum for the transmission coefficient $T$, first for no pump ($\Omega_{\text{pump}}=0$) and for a coupling efficiency $\eta=0.05$ similar to the one measured in the single-atom experiment of Ref. \cite{ref:Kurtsiefer} (Fig. \ref{fig:Transmission}a). We have generalized the above calculations to allow for a non-zero detuning $\Delta=\omega_\text{p}-\omega_{ge}$ between the incoming probe and atomic resonance frequency $\omega_{ge}$ (Eq. (\ref{eq:Transmission_clean}) is obtained for $\Delta=0$). A minimum transmittance on resonance of $T\approx 80\%$ is predicted, as seen in experiments. In Fig. \ref{fig:Transmission}b, we then increase $\Lambda$ to $\{0.25,0.5,1.1\}$ by increasing the pump amplitude. One can observe the complete attenuation of transmission for $\Lambda=1/2$, and also the bump in transmission $T>1$ for values of $\Lambda>1$.
 
\vspace{-0.5em}
 \subsection{Photon number correlations}
Having discussed the effects of the pump beam on the transmission spectra, we continue with the second-order correlation function $g^{(2)}(\tau)$, which characterizes the relative likelihood of detecting two photons separated by a time delay $\tau$ and is defined as
\begin{equation}
    g^{(2)}(\tau)=\frac{\left\langle \hat{E}_{\text{det}}^\dagger(0)\hat{E}_{\text{det}}^\dagger(\tau)\hat{E}_{\text{det}}(\tau)\hat{E}_{\text{det}}(0)\right\rangle}{|\langle \hat{E}_{\text{det}}^\dagger(0)\hat{E}_{\text{det}}(0)\rangle|^2},
    \label{eq:g2_definition}
\end{equation}
where the denominator can be obtained from Eq. (\ref{eq:Transmission_clean}) since $\langle \hat{E}_{\text{det}}^\dagger(0)\hat{E}_{\text{det}}(0)\rangle=T\Phi_\text{p}$. To calculate the numerator of Eq. (\ref{eq:g2_definition}), it is convenient to move to the Schrodinger picture. In that case, the numerator physically describes a process where, starting from the steady state density matrix $\hat{\rho}_ {\text{ss}}$, a photon is detected at time $\tau=0$, which projects the atom into a new conditional state $\hat{\rho}'(0)=\hat{E}_{\text{det}}\hat{\rho}_{\text{ss}}\hat{E}_{\text{det}}^\dagger/\text{Tr}(\hat{E}_{\text{det}}\hat{\rho}_ {\text{ss}}\hat{E}_{\text{det}}^\dagger)$. This state is non-trivial, as the detected photon could have come either from the atom or from the probe beam. Explicitly, the annihilation operator $\hat{E}_{\text{det}}$ (Eq. (\ref{eq:Input_output_projected})) contains contributions both from the probe input and the atomic lowering operator, implying that the resulting atomic state after emission is generally not just $\hat{\rho}'(0)=|g\rangle \langle g|$. The numerator of Eq. (\ref{eq:g2_definition}) then corresponds to the intensity emitted by the system, $\text{Tr}(\hat{E}_{\text{det}}^\dagger\hat{E}_{\text{det}}\hat{\rho}'(\tau))$, as this transient density matrix evolves in time $\tau$ under Eq. (\ref{eq:Master_equation}), eventually returning back to the steady state. It is straightforward to show that
\begin{equation}
    g^{(2)}(\tau)=e^{-\Gamma_0\tau}\left[\left|\frac{2\Lambda}{1-2\Lambda}\right|^2-e^{\Gamma_0\tau/2}\right]^2,
    \label{eq:g2_final}
\end{equation}
within the weak driving approximation. The form of Eq. (\ref{eq:g2_final}) has been previously derived for the case of no pump field (thus $\Lambda=\eta$) \cite{ref:Kurtsiefer,ref:1D_waveguide,ref:Carmichael}. From Eq. (\ref{eq:g2_final}) we see that the form remains unchanged in the presence of a pump, with the coupling efficiency replaced by $\Lambda=\eta\Omega/\Omega_p$. That is, from the standpoint of transmission (Eq. \ref{eq:Transmission_clean}) and also second-order correlation functions, the addition of our pump beam plays \textit{exactly the same role} as achieving a high coupling efficiency, within the weak driving regime and assuming mutually orthogonal pump-probe spatial modes. This constitutes the central result of this work. \\

\begin{figure}[h!]
    \includegraphics[width=1\linewidth]{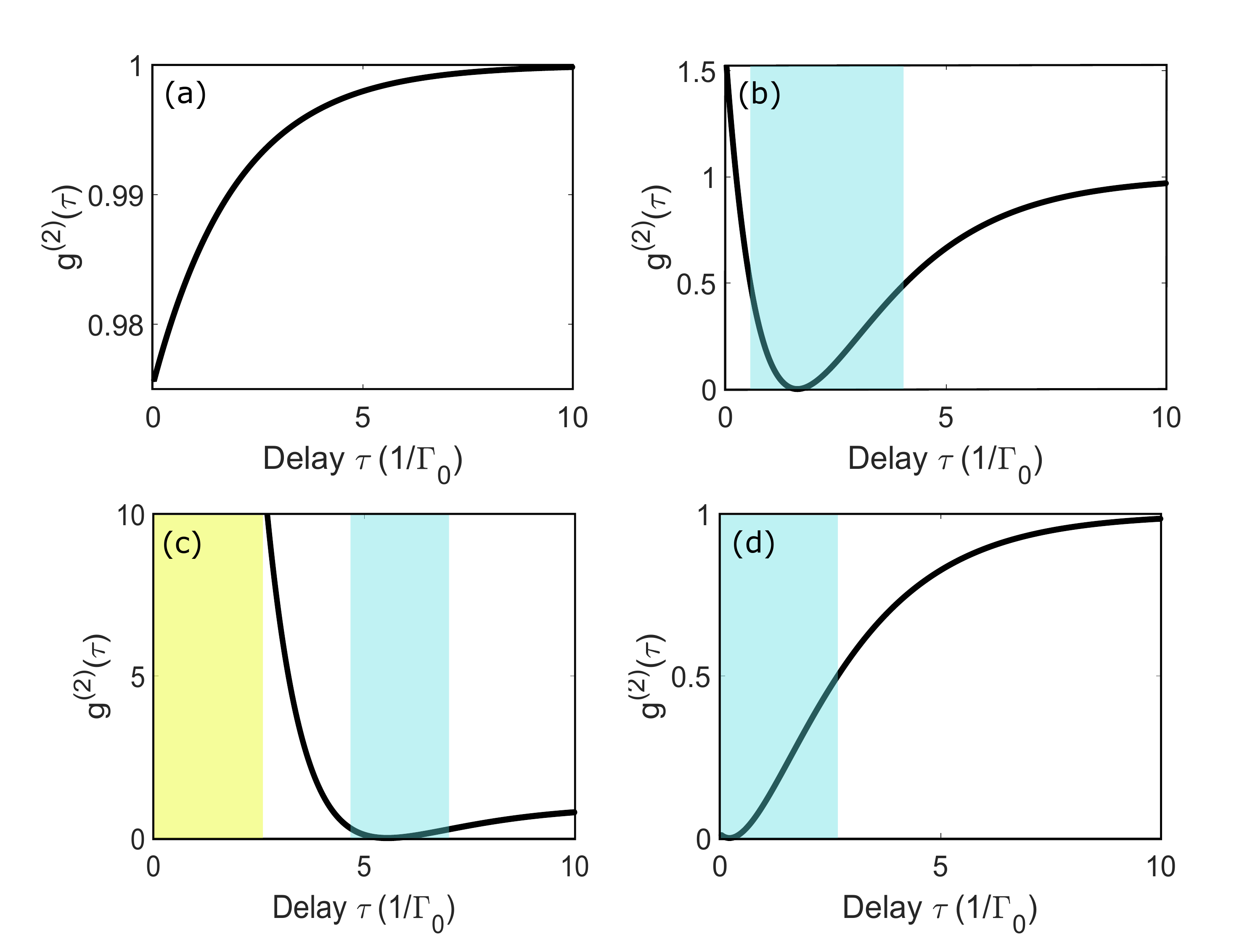}
    \caption{Second-order correlation function $g^{(2)}(\tau)$ of the transmitted field, as a function of time delay $\tau$ for a weak, coherent probe beam. In (a) the pump amplitude is set to zero and we take an experimentally realistic atom-light coupling efficiency of $\eta=0.05$. Due to the low coupling efficiency, the transmitted field mostly consists of the incident coherent field, and thus $g^{(2)}\approx 1$ for any $\tau$. In (b) and (c), a non-zero pump yields an effectively enhanced coupling efficiency of $\Lambda=0.3$ and $\Lambda=0.4$ (respectively), giving rise to non-trivial photon correlations. In (d) we consider the case of $\Lambda=10$, where a pump beam much stronger than the probe causes the total field correlations to be dominated by the atomic scattered field, which is trivially anti-bunched. Light blue regions correspond to notably anti-bunched correlations ($g^{(2)}<0.5$) while light yellow is used to indicate large photon bunching ($g^{(2)}>10$).}
    \label{fig:g2(T)}
\end{figure}

Next, we study the two-photon correlation function within different regimes of interest. We start by considering the case with no pump, where $\Lambda=\eta$. Plotting the result of Eq. ($\ref{eq:g2_final}$) for a low coupling efficiency of $\eta=0.05$, we naturally find that the statistics of the total detected field are dominated by the one from the input coherent state probe, $g^{(2)}(\tau)\approx 1$ (Fig. \ref{fig:g2(T)}a). On the other hand, by tuning the pump such that $\Lambda=1/2$, one can achieve extreme bunching at zero time delay, $g^{(2)}(0)\rightarrow\infty$. More precisely, an exact calculation shows that $g^{(2)}(0)\propto(\Gamma_0/\Omega)^{4}$, as the Rabi frequency is reduced. The large bunching coincides with the suppression of the linear transmission $T=0$ in the denominator of Eq. (\ref{eq:g2_definition}). Physically, with the complete cancellation of the linear response, the remaining photons detected in $g^{(2)}$ are those arising from nonlinear processes, where the atom acts as a frequency mixer and the re-scattered photons propagate past the atom in a correlated fashion \cite{ref:quantum_impurity,ref:generalized_input_output}. Furthermore, in cases where strong bunching is observed at $\tau=0$, a perfect anti-bunching $g^{(2)}(\tau)=0$ of the photon correlations can be found at later delay times $\tau_{A}=(4/\Gamma_0)\ \text{ln}\left|2\Lambda/(2\Lambda-1)\right|$ (Fig. \ref{fig:g2(T)}b,c). The expression for $\tau_A$ also holds for the parameter regime $1/4\leq \Lambda < 1/2$, and in particular, for $\Lambda=1/4$, one finds perfect anti-bunching at $\tau=\tau_A=0$. As we discuss later, this can be understood from the transient atomic state following the detection of the first photon, which happens to instantaneously emit light with an amplitude and phase that cancels the probe beam, before relaxing back to equilibrium. Finally, for large values of $\Lambda\gg1$, the scattered field dominates over the probe, giving rise to the characteristic anti-bunching of the pure atomic emission (Fig. \ref{fig:g2(T)}d).

\begin{figure}[h!]
  \begin{center}
    \includegraphics[width=1\linewidth]{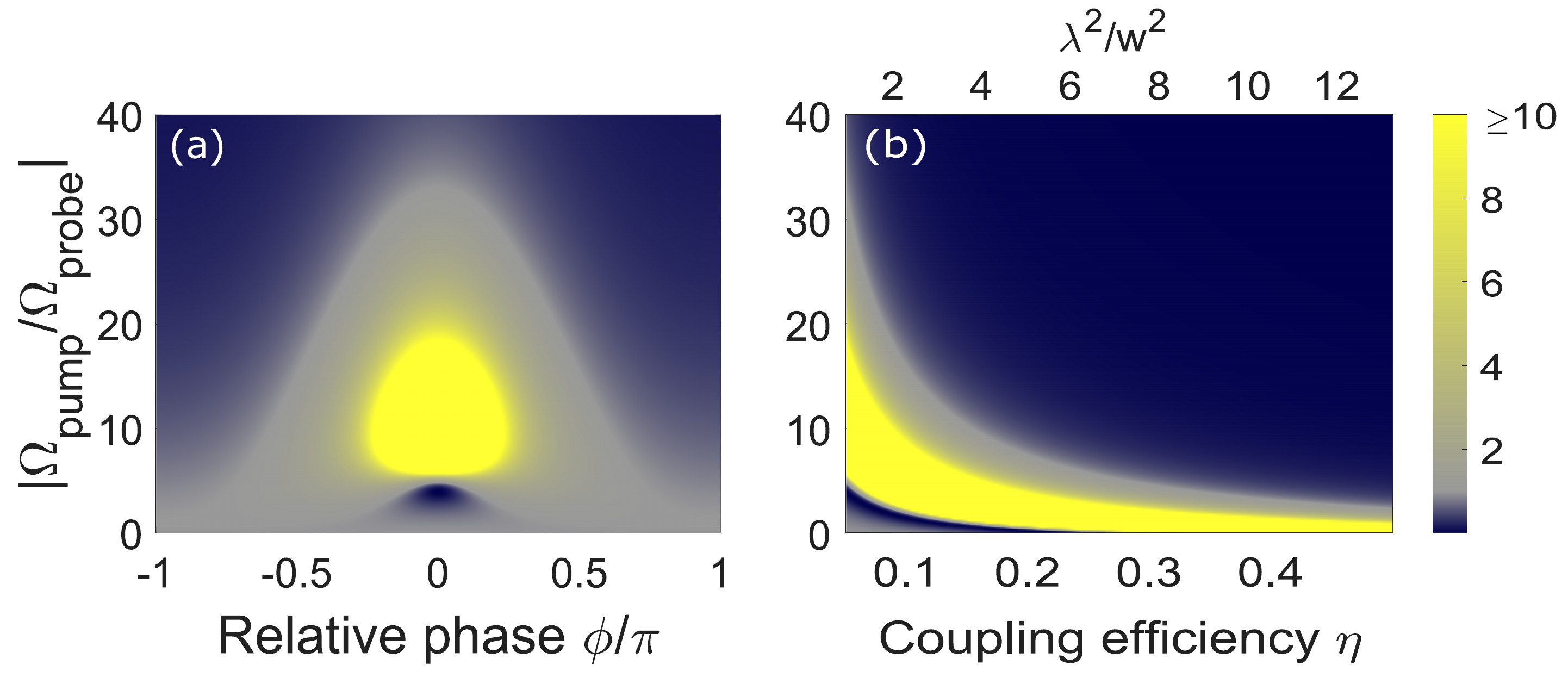}
    \caption{$g^{(2)}(0)$ as a function of the coupling efficiency $\eta$, the pump/probe relative phase $\phi$ and the amplitude ratio $|\Omega_{\text{pump}}/\Omega_{\text{probe}}|$. In (a) we fix $\eta=0.05$ and plot $g^{(2)}(0)$ as a function of $|\Omega_{\text{pump}}/\Omega_{\text{probe}}|$ and $\phi$. In (b) we show again the values for $g^{(2)}(0)$ but now fixing $\phi=0$ and exploring different $\eta$. Additionally, we show the beam waist $w$ associated to the values of $\eta$ assuming a spatial Gaussian detection mode within the paraxial approximation. As extreme values of bunching are possible, we represent all the values $g^{(2)}(0)\geq 10$ by the same color. }
    \label{fig:g2(0)}
  \end{center}
\end{figure}

Overall, we see that just with a proper tuning of the pump field, one can switch from completely anti-bunched to extremely bunched photon correlations in the total transmitted field. While thus far we considered the relative phase of the pump beam $\phi=0$ to be fixed, we can alternatively plot the second-order correlation $g^{(2)}(0)$ for varying amplitude $|\Omega_{\text{pump}}|$ and relative phase, as shown in Fig. \ref{fig:g2(0)}a (for a fixed efficiency $\eta=0.05$). It can be seen that the strong anti-bunching and bunching features exist within a reasonable tolerance of the relative phases and amplitudes. Likewise, in Fig. \ref{fig:g2(0)}b, we consider a fixed relative phase $\phi=0$ and vary the efficiency $\eta$. For reference, we also provide the beam waist $w$ corresponding to this coupling efficiency, assuming a paraxial, Gaussian detection mode (where $\eta=3\lambda^2/8\pi^2w^2$).

\vspace{-3em}
\section{Connection with the transient atomic state}
In this section, we present an alternative description for the origin of the second-order correlations, now from the atomic state perspective. As shown in the previous section, starting from the steady state, the detection of a photon at $t=0$ projects the atom into the new conditional state $\hat{\rho}'(0)$ (which in general is different from the ground state). With this, $g^{(2)}(\tau)$ is then related to the average intensity of the conditioned total transmitted field since $g^{(2)}(\tau)=\langle \hat{E}_{\text{det}}^\dagger \hat{E}_{\text{det}}\rangle_{\hat{\rho}'(\tau)}/T$, with $\hat{\rho}'(\tau)$ obeying Eq. (\ref{eq:Master_equation}). It is helpful to explicitly write out the solution to the transient atomic coherence, 
\begin{equation}
  \hat{\rho}_{ge}'(\tau)=\frac{-2i\Omega}{\Gamma_0}\left(1+\frac{2\Lambda}{1-2\Lambda}e^{-\Gamma_0\tau/2}\right),
  \label{eq:coherence}
\end{equation}
where we again neglect terms of order $\Omega^2$. The population in the excited state can be obtained through $\hat{\rho}_{ee}'(t)=|\hat{\rho}_{ge}'(t)|^2$ within the weak driving approximation. From Eq. (\ref{eq:coherence}), the atomic coherence of the projected state right after the detection of a photon reads $\hat{\rho}_{ge}'(0)=-2i\Omega/[\Gamma_0(1-2\Lambda)]$. Then, when there is no pump and $\eta$ is small, one can see that $\hat{\rho}_{ge}'(0)\approx \hat{\rho}_{ge}^{\text{ss}}$ as the first photon measurement barely affects the atom. For $\Lambda=1/4$ (perfect anti-bunching), the atom in the conditional state radiates a scattered field that cancels the input probe amplitude at $\tau=0$, as can be confirmed by substituting the conditional atomic coherence $\hat{\rho}_{ge}'(0)$ into Eq. (\ref{eq:Input_output_projected}). If $\Lambda>1/4$, the conditional total field has the opposite sign at $\tau=0$ compared to its steady state value. Thus, as it relaxes back to equilibrium, $g^{(2)}(\tau)$ becomes anti-bunched at $\tau=\tau_A$, when the conditional total field switches sign and passes through zero.

Exactly at $\Lambda=1/2$, individual photons cannot be transmitted through the atom when the system is in its steady state, as the linear transmission $T=0$ from Eq. (\ref{eq:Transmission_clean}). Instead, the only transmission events for weak driving consist of photon pairs, which are frequency mixed by the atom. The individual photons of this pair have no well-defined phase, but are frequency correlated with each other \cite{ref:quantum_impurity,ref:generalized_input_output}. This lack of phase is reflected in the conditional atomic density matrix. In this case, the linear approximation $\hat{\rho}_{ge}'(0)=-2i\Omega/[\Gamma_0(1-2\Lambda)]$ breaks down, and an exact calculation reveals that the atom is completely mixed, with $\hat{\rho}_{ee}'(0)=\hat{\rho}_{gg}'(0)=1/2$.

\section{Conclusions}

We have described a new scenario in which a single, laser-illuminated atom can produce strongly non-classical photon correlations in the total field. Considering two laser beams that meet at the atom, we find scenarios in which the detected atomic contribution can easily be made comparable in strength to the detected laser contribution. We show that the effect of the second beam, which acts as a pump, is in some scenarios formally equivalent to an increased atom-light coupling efficiency. This allows one to achieve correlations associated with coupling efficiencies beyond what is practically possible, and even beyond what is physically possible, i.e., above 100\%. We expect that our method can be immediately applied to observe interesting quantum behaviour in existing experiments where single atoms are coupled to tightly focused beams. It would also be interesting in the future to explore more broadly whether other single-atom, quantum scattering phenomena can be effectively ``amplified'' by similar techniques.

\section*{Acknowledgements}
We thank S. Grava and H. J. Kimble for valuable discussions. The authors acknowledge funding from Spanish MINECO projects OCARINA (Grant Ref. PGC2018-097056-B-I00), Q-CLOCKS (PCI2018-092973) and MINECO Severo Ochoa Grant SEV-2015-0522; Plan Nacional Grant ALIQS, funded by MCIU, AEI, and FEDER; Fundación Ramón Areces Project CODEC; Fundació Mir-Puig; Fundacio Privada Cellex; Ag\`{e}ncia de Gesti\'{o} d'Ajuts Universitaris i de Recerca (AGAUR) project (2017-SGR-1354);  Generalitat de Catalunya (CERCA Programme, RIS3CAT project QuantumCAT); the Secretaria d’Universitats i Recerca de la Generalitat de Catalunya and the European Social Fund (2020 FI B 00196); ERC Starting Grant FOQAL; Quantum Technology Flagship projects MACQSIMAL (820393) and QRANGE (820405); and 17FUN03-USOQS, which has received funding from the EMPIR programme co-financed by the Participating States and from the European Union's Horizon 2020 research and innovation programme.

\bibliographystyle{apsrev4-1} 

\bibliography{bibliography}

\clearpage
\subsection*{Appendix: Derivation of the mode-projected operators}
\vspace{-0.5em}
Here we derive the projection of $\hat{\textbf{E}}_{\text{out}}(\textbf{r})$, Eq. (\ref{eq:Input_output_projected}) in the main text, starting from Eq. (\ref{eq:Input_output_not_projected}), following closely the arguments in Ref. \cite{ref:Marcos_paper}. The quantized electromagnetic field can be expressed as a combination of plane-wave operators of the form $\hat{\textbf{E}}_{\textbf{k},\hat{\epsilon}_{\textbf{k},j}}(\textbf{r})=E_0(k)\textbf{u}_{\textbf{k},\hat{\epsilon}_{\textbf{k},j}}(\textbf{r})\hat{a}_{\textbf{k},\hat{\epsilon}_{\textbf{k},j}}$, labelled by the wave-vector $\textbf{k}$ and the orthonormal polarizations $\hat{\epsilon}_{\textbf{k},j}$ with $j=\{1,2\}$ and $\textbf{k}\cdot \hat{\epsilon}_{\textbf{k},j}=0$. Here, we define $\textbf{u}_{\textbf{k},\hat{\epsilon}_{\textbf{k},j}}(\textbf{r})= e^{-i \textbf{k}\cdot \textbf{r}}\hat{\epsilon}_{\textbf{k},j}$ as the plane-wave spatial mode, $\hat{a}_{\textbf{k},\hat{\epsilon}_{\textbf{k},j}}$ is the associated photon annihilation operator, and $E_0(k)$ is a normalization factor, whose specific form is not relevant here. For a fixed $|\textbf{k}|=k_0$, one can alternatively construct a field operator $\hat{\textbf{E}}_\alpha (\textbf{r})$ based on any superposition of plane waves of the same $|\textbf{k}|$, 
\begin{equation}
    \hat{\textbf{E}}_\alpha (\textbf{r})=E_0(k_0)\hat{a}_\alpha \textbf{E}_{\alpha}(\textbf{r}),
\end{equation}
where the associated spatial mode reads
\begin{multline}
    \textbf{E}_\alpha(\textbf{r})=\frac{k_0^2}{(2\pi)^2}\sum\limits_{j=1,2}\int_0^{2\pi}\text{d}\phi \int_{0}^{\pi}\sin\theta\ \text{d}\theta\   c_{\alpha,\theta,\phi,j}\times \\ \times e^{-ik_0(x\sin\theta\cos\phi+y\sin\theta\sin\phi+z\cos\theta)}\hat{\epsilon}_{k_0,j}.
    \label{eq:Field_decomposition}
\end{multline}
Here, we have utilized the spherical coordinates $k_x=k_0\sin{\theta}\cos{\phi}$, $k_y=k_0\sin{\theta}\sin{\phi}$, $k_z=k_0\cos{\theta}$ and d$V_k=k_0^2\sin{\theta}\text{d}\theta\text{d}\phi$ to express the linear combination as an integral over solid angle. The scalar product (mode overlap) between two arbitrary spatial modes $\textbf{E}_\alpha(\textbf{r})$ and $\textbf{E}_\beta(\textbf{r})$ is defined as the two-dimensional integral over any fixed plane $z=\text{const.}$
\begin{equation}
    \langle \textbf{E}_\alpha|\textbf{E}_\beta\rangle\equiv\iint\limits_{z=\text{const.}} \text{d}^2\textbf{r}\ \textbf{E}_\alpha^*(\textbf{r})\cdot \textbf{E}_\beta(\textbf{r}),
    \label{eq:scalar_product_zcte}
\end{equation}
where the plane wave modes fulfill the orthonormality relation
\begin{equation}
\langle \textbf{u}_{k_0,\phi',\theta',j'}|\textbf{u}_{k_0,\phi,\theta,j}\rangle=\frac{(2\pi)^2\delta^{jj'}}{k_0^2\text{sin}\theta}\delta(\theta-\theta')\delta(\phi-\phi').
\label{eq:normalization_overlap}
\end{equation} 
The spatial mode overlap from Eq. (\ref{eq:scalar_product_zcte}) can be related to the electromagnetic field power $P$ defined as the integral of the z-component of the Poynting vector in the plane $z=0$. Explicitly,
\begin{equation}
    P_\alpha=2\epsilon_0c\int_{z=0}\text{d}^2 \textbf{r}\ \textbf{E}_\alpha^*(\textbf{r})\cdot  \textbf{E}_\alpha(\textbf{r})=2\epsilon_0c\ \langle \textbf{E}_\alpha|\textbf{E}_\alpha\rangle.
    \label{eq:Normalization_fields}
\end{equation}
From here, the field operators can be conveniently re-normalized such that expectation values of the form $\langle \hat{E}_{\alpha}^\dagger\hat{E}_{\alpha}\rangle$ are in units of photons per unit time. To do so, we start by considering a field $\hat{E}_{\text{det}}(\textbf{r})$ in the spatial detection mode $\textbf{E}_{\text{det}}(\textbf{r})$. Multiplying the fields by the normalization constant $N$, we impose that $\langle N\hat{E}_{det}^\dagger N\hat{E}_{det}\rangle\equiv P_{\text{det}}/\hbar\omega_{ge}$ yielding $N=\sqrt{2\epsilon_0/\hbar k_0F_{\text{det}}}$, where $F_{\text{det}}=|\langle \textbf{E}_{\text{det}}|\textbf{E}_{\text{det}}\rangle|$. This normalization will be implicit for the rest of this section.\\

Next, we apply the previous ideas to evaluate the projection of the atomic scattered field operator $\hat{\textbf{E}}_\text{sc}(\textbf{r})$ (second term in Eq. \ref{eq:Input_output_not_projected}) into any desired detection mode $\textbf{E}_{\text{det}}(\textbf{r})$, using Eq. (\ref{eq:Field_decomposition}) and Eq. (\ref{eq:scalar_product_zcte}), by writing both the Green's function and detection mode in a plane wave expansion \cite{ref:Marcos_paper}. This gives, within the previous normalization, 
\begin{equation}
    \langle \textbf{E}_{\text{det}}|\hat{\textbf{E}}_{\text{sc}}\rangle =
   id_{eg}\sqrt{ \frac{k_0}{2\hbar\epsilon_0F_{\text{det}}}}\textbf{E}_{\text{det}}^*(\textbf{r}_\text{a})\cdot \textbf{d}\ \hat{\sigma}^{ge},
    \label{eq:scattered_field_overlap}
\end{equation}
where $\textbf{E}_{\text{det}}^*(\textbf{r}_\text{a})$ is the conjugate of the amplitude of the spatial detection mode evaluated at the atomic position $\textbf{r}_a$. We note that although the mode function $\textbf{E}_{\text{det}}(\textbf{r})$ can be arbitrarily rescaled by a global coefficient, this freedom is eliminated in Eq. (\ref{eq:scattered_field_overlap}) through the normalization constant $F_{\text{det}}$, thus making the result of the overlap clearly defined. \\

Now, we will show that Eq. (\ref{eq:scattered_field_overlap}) can be written in an even simpler way, involving only the collection efficiency $\eta$ defined as the probability that a photon emitted by the atom is measured in the detection mode. For convenience, we consider the scenario where the atom starts in the excited state and it is not driven by any field, such that $\hat{\rho}^{ee}(t)=e^{-\Gamma_0t}$ and a single photon is emitted as $t\rightarrow\infty$. The explicit value of $\eta$ is then given by the time integral of the overlap from Eq. (\ref{eq:scattered_field_overlap}),
\begin{equation}
    \eta=\int_0^\infty\text{d}t|\langle \textbf{E}_{\text{det}}|\hat{\textbf{E}}_{\text{sc}}(t)\rangle|^2=\frac{3\pi}{2k_0^2}\frac{|\textbf{E}_{\text{det}}(\textbf{r}_\text{a})\cdot \textbf{d}|^2}{F_{\text{det}}}.
    \label{eq:eta}
\end{equation}
For a detection mode that matches exactly the radiation pattern of a point dipole over $4\pi$ (all solid angle), one obtains a maximum of $\eta=1$. We notice that Eq. (\ref{eq:eta}) allows us to establish the relation $\sqrt{\Phi_\text{p}}=|\Omega_{\text{probe}}|/\sqrt{\eta\Gamma_0}$ from the main text, as $\Phi_\text{p}=2\epsilon_0c|\langle \textbf{E}_{\text{p}}|\textbf{E}_{\text{p}}\rangle|/\hbar\omega_{ge}$ for the probe beam in the detection mode. Substituting the collection efficiency $\eta$ from Eq. (\ref{eq:eta}) into  Eq. (\ref{eq:scattered_field_overlap}), we arrive at Eq. (\ref{eq:Input_output_projected}) in the main text.\\
 
 Finally, we explicitly calculate the collection efficiency from Eq. (\ref{eq:eta}) for the particular case of a spatial, Gaussian input (and detection) mode, within the paraxial approximation. As commented in the main text, the spatial mode takes the form $E_{\text{p}}(\rho,z=0)=E_0e^{-\rho^2/w^2}$ which implies that $ \langle \textbf{E}_{\text{p}}|\textbf{E}_{\text{p}}\rangle=E_0^2\pi w^2$. Thus, assuming that the field is polarized along $\textbf{d}$, the coupling efficiency takes the value $\eta=3\lambda^2/8\pi^2w^2$, as in the main text. 
\end{document}